\documentclass[twocolumn,pra,aps,showpacs]{revtex4}
\usepackage{amssymb}
\usepackage{amsmath}
\usepackage{epsfig}

\begin{document}

\title{Self-Trapping of Bose-Einstein Condensates in an Optical Lattice:
the Effect of the System Dimension }
\author{ Ju-Kui Xue$^{1,2}$}
\author{ Ai-Xia Zhang$^{1}$ }
\author{ Jie Liu$^{2,}$ }
\email[Email: ]{liu_jie@iapcm.ac.cn}

\affiliation{$^{1}$Physics and Electronics Engineering College, Northwest Normal
University, 730070 Lanzhou, People's Republic of China\\
$^{2}$Institute of Applied Physics and Computational Mathematics, P.O. Box
8009 (28), 100088 Beijing, People's Republic of China}

\begin{abstract}
In the present paper, we investigate the dynamics  of a
Bose-Einstein condensates (BEC) loaded into an deep optical lattice
of 1D, 2D and 3D,  both analytically and numerically. We focus on
the self-trapping state and the effect of the system dimension.
Under the tight-binding approximation we obtain an analytical
criterion for the self-trapping state of BEC  using time-dependent
variational method.  The phase diagram for self-trapping, soliton,
breather, or diffusion of the BEC cloud is obtained accordingly and
verified by directly solving the discrete Gross-Pitaevskii equation
(GPE) numerically. In particular, we find that the criterion and the
phase diagrams are  modified dramatically by the dimension of the
lattices.
\end{abstract}

\maketitle

\section{Introduction}
Recently, the Bose-Einstein condensates (BECs) in an optical lattice
attract much attention both experimentally and
theoretically\cite{rev1}. This is partly  because the physical
parameters, e.g., lattice parameters, interaction strength, etc., in
such a  clean  system ("clean" means perfect periodicity without
impurity) can be manipulated at will using modern experimental
technique like Feshbach resonance. Taking this advantage  people
have observed well-known and long-predicted phenomena, such as Bloch
oscillations\cite{d1} and the quantum phase transition between
superfluid and Mott insulator\cite{d2}. More importantly, there are
new phenomena that have been either observed or predicted in this
system, for example, nonlinear Landau-Zener tunneling between Bloch
bands\cite{d3,d4} and the strongly inhibited transport of
one-dimensional BEC in an optical lattice\cite{d5}.

Another intriguing phenomenon, namely, self-trapping, was recently
observed experimentally in this system \cite{c10}. In such a
experiment, it was observed that a BEC cloud with repulsive
interaction initially loaded in an optical lattice was self-trapped
or localized and the diffusion was completely  blocked when the
number of the degenerate bosonic atom was larger than a critical
value. This is quite counterintuitive. Even without interaction, a
wave packet with a narrow distribution in the Brillouin zone expands
continuously inside a periodic potential. When there is an
interaction between atoms and it is repulsive, one would certainly
expect the wave packet to expand.  On the other aspect, the
self-trapping phenomenon also attract many theoretical efforts and a
lot of meaningful results have been
obtained\cite{c11,c12,c13,c14,c15,kivshar,c16}. However, most of
these analyses are limited to 1D system. Beyond these 1D
approximation,
 the localized states\cite%
{c17}-\cite{c18} , self-trapping and other quantum phenomena\cite{c19}-\cite%
{c22} have been observed in high-dimensional optical lattices. To
our knowledge, however, there are no systematical theoretical
analyses on the dynamics of BEC loaded into multi-dimensional
optical lattices. There leaves some  important problems, e.g., what
is the effect of the system dimension on the dynamics of BEC in
optical lattices ?

In this paper, we address this issue both analytically and
numerically  focusing on the intriguing self-trapping phenomenon,
within  the tight-binding approximation. Based on the discrete GPE
and exploiting a time-dependent variational approach, we obtain an
analytical criterion for the self-trapping state of BEC and   the
phase diagram for self-trapping, soliton, breather, or diffusion of
the BEC cloud. Our results show that, the change on the dimension of
the system leads to many interesting consequence: i) Self-trapping
state can also emerge in higher-dimension systems, however, compared
to the 1D situation its parameter domain in the phase diagram is
largely shrunken; ii) The stable moving soliton and breather
solutions of the wavepackets that exist in $1D$ system, no longer
stand  in $2D$ and $3D$ systems. iii) In addition, when the
self-trapping occurs, the ratio between  the  final BEC cloud width
and the initial wavepacket width is greatly modified in $2D$ and
$3D$ systems compared to the 1D case.

Our paper is organized as follows. In Sec.II, under the
tight-binding approximation, we obtain a discrete GPE that governs
the dynamical behavior of BEC in deep $3D$ periodic optical
lattices. In Sec.III, using the time-dependent variation of a
Gaussian profile wavepacket, the dynamics of BEC  is discussed
analytically and the critical conditions for occurrence of
self-trapping state are obtained. Moreover, the phase diagram for
self-trapping, soliton, breather, or diffusion of the BEC matter
waves is obtained accordingly. In Sec.IV,  directly solving  GPE is
exploited to confirm our theoretical predictions. Sec.V, are our
discussion and conclusion.

\section{NONLINEAR DISCRETE MODEL}

We focus our attention on the situation that a BEC is loaded into a $3D$
optical lattices. In the mean field approximation, the $3D$ BEC dynamics at
T=0 satisfies the GPE%
\begin{equation}
i\hslash \frac{\partial \Psi }{\partial t}=-\frac{\hslash ^{2}}{2M}\nabla
^{2}\Psi +(V_{opt}(\overrightarrow{r})+\frac{4\pi \hslash ^{2}a_{s}}{M}|\Psi
|^{2})\Psi  \label{1}
\end{equation}%
where $\Psi $ is the wave function of the condensate. The coupling constant $%
4\pi \hslash ^{2}a_{s}/M$, which relate to the mass of the atoms $M$ and the
repulsive (attraction) s-wave scattering length $a_{s}>0(a_{s}<0)$, is the
nonlinear coefficient corresponding to the two body interactions, $V_{opt}(%
\overrightarrow{r})=V_{\max }(\cos (2\kappa x)+$ $\cos (2\kappa y)+\cos
(2\kappa z))$ is the external optical potential created by three orthogonal
pairs of counter-propagating laser beams\cite{c23}, $\kappa $ is the wave
number of the laser lights that generate the optical lattice. $\bigskip $All
the variables can be re-scaled to be dimensionless by the system's basic
parameters $t$ $\sim $ $t\nu $, $x$ $\sim $ $\kappa x$, $\Psi \sim $ $\frac{1%
}{\sqrt{\kappa N}}\Psi $, then GPE (1) becomes

\begin{equation}
i\frac{\partial \Psi }{\partial t}=-\nabla ^{2}\Psi +V(\overrightarrow{r}%
)\Psi +\frac{4\pi \hslash ^{2}a_{s}\kappa N}{M\hslash \nu }|\Psi |^{2}\Psi
\label{2}
\end{equation}%
where $N$ is the total number of atoms, $V(\overrightarrow{r})=V_{0}[\cos
(2x)+$ $\cos (2y)+\cos (2z)]$, the parameter $V_{0}=V_{\max }/2E_{R}$ is the
strength of the optical lattices, with $E_{R}=\hslash ^{2}\kappa ^{2}/2M$ is
the recoil energy of the lattice, $\nu =E_{R}/\hslash $. The associated
Hamiltonian is

\begin{equation}
H=\int d\overrightarrow{r}[-\Psi (\overrightarrow{\nabla }\Psi )\Psi ^{\ast
}+V(\overrightarrow{r})|\Psi |^{2}+\frac{2\pi \hslash ^{2}a_{s}\kappa N}{%
M\hslash \nu }|\Psi |^{4}]  \label{3}
\end{equation}

In order to understand the dynamical behaviors of BEC in deep 3D optical
lattices, it is worthwhile to study the system in certain extreme limits.
The particular limit that we now insight into is the tight-binding
approximation, where the lattice is so strong that the BEC system can be
looked as a chain of trapped BECs that are weakly linked\cite{c11}. Under
the tight binding limit, the condensate parameter can be written as

\begin{equation}
\Psi (\overrightarrow{r},t)=\sum_{n,m,k}\psi _{n,m,k}(t)\phi (%
\overrightarrow{r}-\overrightarrow{r}_{n,m,k})  \label{4}
\end{equation}

where $\psi _{n,m,k}(t)=\sqrt{N_{n,m,k}(t)}e^{i\theta _{n,m,k}(t)}$ is the ($%
n,m,k)$th amplitude, $N_{n,m,k}$ and $\theta _{n,m,k}(t)$ are the number of
particles and phases in the well $(n,m,k)$. $\sum_{n,m,k}|\Psi
_{n,m,k}|^{2}=1$, $\phi (\overrightarrow{r}-\overrightarrow{r}_{n,m,k})$ is
the condensate wavefunction localized in the well $(n,m,k)$, with the
normalization $\int d\overrightarrow{r}\phi _{n,m,k}^{2}=1$, and the
orthogonality of the condensate wave functions $\int d\overrightarrow{r}\phi
_{n,m,k}\phi _{n+1,m,k}\simeq 0,\int d\overrightarrow{r}\phi _{n,m,k}\phi
_{n,m+1,k}\simeq 0,\int d\overrightarrow{r}\phi _{n,m,k}\phi
_{n,m,k+1}\simeq 0$. Inserting the nonlinear tight-binding approximation (4)
into the GPE (2) and integrating out the spatial degrees of freedom,
following re-scale the time as $t\rightarrow 2Kt$, we find the reduced GPE

\begin{eqnarray}
i\frac{\partial \psi _{n,m,k}}{\partial t} &=&-\frac{1}{2}(\psi
_{n-1,m,k}+\psi _{n+1,m,k}+\psi _{n,m-1,k}  \notag \\
&&+\psi _{n,m+1,k}+\psi _{n,m,k-1}+\psi _{n,m,k+1})  \notag \\
&&+(A+\Lambda |\psi _{n,m,k}|^{2})\psi _{n,m,k}  \label{5}
\end{eqnarray}

where $A=\frac{1}{2K}\int d\overrightarrow{r}[(\nabla \phi _{n,m,k})^{2}+V(%
\overrightarrow{r})\phi _{n,m,k}{}^{2}]$, $\Lambda =\frac{1}{2K}\frac{4\pi
\hslash ^{2}a_{s}\kappa N}{M\hslash \nu }\int d\overrightarrow{r}\phi
_{n,m,k}{}^{4}$ is the ratio of the nonlinear coefficient, induced by the
two-body interatomic interactions. And the included parameter $K\simeq -\int
dr\{(\nabla \phi _{n+1,m,k}\nabla \phi _{n,m,k}+\nabla \phi _{n,m+1,k}\nabla
\phi _{n,m,k}+\nabla \phi _{n,m,k+1}\nabla \phi _{n,m,k})+V_{0}[\cos
(2x)\phi _{n+1,m,k}\phi _{n,m,k}+\cos (2y)\phi _{n,m+1,k}\phi _{n,m,k}+\cos
(2z)\phi _{n,m,k+1}\phi _{n,m,k}]\}$ is the tunneling rates between the
adjacent sites $(n,m,k)$ and $(n+1,m+1,k+1)$. In order to make an agreement
with the experiments done with $^{87}Rb$, the two-body interaction term is
expected to be positive, that is $\Lambda >0$.

Equation (5) can be written as $\overset{\cdot }{\psi }_{n,m,k}=\partial
H/\partial (i\psi _{n,m,k}^{\ast })$, where $H$ is the Hamiltonian function

\begin{eqnarray}
H &=&\sum_{n,m,k}[-\frac{1}{2}(\psi _{n,m,k}\psi _{n+1,m,k}^{\ast }+\psi
_{n,m,k}^{\ast }\psi _{n+1,m,k}  \notag \\
&&+\psi _{n,m,k}\psi _{n,m+1,k}^{\ast }+\psi _{n,m,k}^{\ast }\psi _{n,m+1,k}
\notag \\
&&+\psi _{n,m,k}\psi _{n,m,k+1}^{\ast }+\psi _{n,m,k}^{\ast }\psi _{n,m,k+1})
\notag \\
&&+A|\psi _{n,m,k}|^{2}+\frac{\Lambda }{2}|\psi _{n,m,k}|^{4}]  \label{6}
\end{eqnarray}

\bigskip

\section{ANALYTIC RESULTS BASED ON THE TIME-DEPENDENT VARIATIONAL APPROACH}

To study the dynamics of the BEC wavepacket loaded into multi-dimensional
optical lattices, we consider the evolution of a Gaussian profile
wavepacket, which we parametrize as $\psi _{n,m,k}(t)=\sqrt{\rho }\exp \{-%
\frac{(n-\xi _{x})^{2}}{R_{x}^{2}}-\frac{(m-\xi _{y})^{2}}{R_{y}^{2}}-\frac{%
(k-\xi _{z})^{2}}{R_{z}^{2}}+i[p_{x}(n-\xi _{x})+p_{y}(m-\xi
_{y})+p_{z}(k-\xi _{z})]+\frac{i}{2}[\delta _{x}(n-\xi _{x})^{2}+\delta
_{y}(m-\xi _{y})^{2}+\delta _{z}(k-\xi _{z})^{2}\}$, define $\alpha =x,y$ or
$z$, where $\xi _{\alpha }(t)$ and $R_{\alpha }(t)$ are the center and the
width of the wavepacket in the $\alpha $ direction, $p_{\alpha }(t)$ and $%
\delta _{\alpha }(t)$ are their associated momenta in the $\alpha $
direction, and $\rho $ is a normalization factor. The wavepacket dynamical
evolution can be obtained by a variational principle from the Lagrangian $%
L=\sum_{n,m,k}i(\overset{\cdot }{\psi }_{n,m,k}\psi _{n,m,k}^{\ast }-\psi
_{n,m,k}^{\ast }\overset{\cdot }{\psi }_{n,m,k}^{\ast })-H$, with the
equations of motion for the variational parameters $q_{i}(t)=$ $\xi _{\alpha
},R_{\alpha },p_{\alpha },\delta _{\alpha }$ given by $\frac{d}{dt}\frac{%
\partial L}{\partial \overset{\cdot }{q_{i}}}=\frac{\partial L}{\partial
q_{i}}$, where $H$ is given by Eq.(6). In calculation, we can replace the
sums over $(n,m,k)$ in the Lagrangian with integrals when $R_{\alpha }$ is
not too small\cite{c11}. In this limit, the normalization factor is obtained
as $\rho =\frac{1}{\sqrt{\pi /2}^{3}\underset{\alpha =x,y,z}{\prod }%
R_{\alpha }}$ from $\int_{-\infty }^{\infty }\int_{-\infty }^{\infty
}\int_{-\infty }^{\infty }|\psi _{n,m,k}|^{2}dndmdk=1$. All these parameters
lead to $L=-V(R_{_{\alpha }},\xi _{_{\alpha }})-\frac{\Lambda }{2\sqrt{\pi }%
^{3}\underset{\beta =x,y,z}{\prod }R_{\beta }}+\sum_{_{_{\alpha
}}}(2p_{_{\alpha }}\overset{\cdot }{\xi }_{_{\alpha }}-\frac{1}{4}\overset{%
\cdot }{\delta }_{_{\alpha }}R_{_{\alpha }}^{2}+e^{-\sigma _{\alpha }}\cos
p_{_{\alpha }})$, where $\sigma _{\alpha }=\frac{1}{2R_{\alpha }^{2}}+\frac{%
R_{\alpha }^{2}\delta _{\alpha }^{2}}{8}$, and $V(R_{_{\alpha }},\xi
_{_{\alpha }})=\rho \int_{-\infty }^{\infty }\int_{-\infty }^{\infty
}\int_{-\infty }^{\infty }dndmdkAe^{-(\frac{2(n-\xi _{x})^{2}}{R_{x}^{2}}+%
\frac{2(m-\xi _{y})^{2}}{R_{y}^{2}}+\frac{2(k-\xi _{z})^{2}}{R_{z}^{2}})}$.
The equations of motion of collective variables $\xi _{\alpha }(t),$ $%
R_{\alpha }(t),$ $p_{\alpha }(t),$ $\delta _{\alpha }(t) $ are given by

\begin{eqnarray}
\overset{\cdot }{p_{\alpha }} &=&-\frac{1}{2}\frac{\partial V}{\partial \xi
_{\alpha }}  \notag \\
\overset{\cdot }{\xi _{\alpha }} &=&\frac{1}{2}e^{-\sigma _{\alpha }}\sin
p_{\alpha }  \notag \\
\overset{\cdot }{\delta _{\alpha }} &=&\cos p_{\alpha }(\frac{2}{R_{\alpha
}^{4}}-\frac{\delta _{\alpha }^{2}}{2})e^{-\sigma _{\alpha }}-\frac{2}{%
R_{\alpha }}\frac{\partial V}{\partial R_{\alpha }}  \notag \\
&&+\frac{\Lambda }{\sqrt{\pi }^{3}R_{\alpha }^{3}\underset{\underset{(\beta
\neq \alpha )}{\beta =x,y,z}}{\prod }R_{\beta }}  \notag \\
\overset{\cdot }{R_{\alpha }} &=&\frac{R_{\alpha }}{2}\delta _{\alpha
}e^{-\sigma _{\alpha }}\cos p_{\alpha }  \label{7}
\end{eqnarray}

\bigskip For simplicity, we assume: $\ R_{x}=R_{y}=R_{z}=R,$ $%
p_{x}=p_{y}=p_{z}=p,$ $\delta _{x}=\delta _{y}=\delta _{z}=\delta ,$ $\sigma
_{x}=\sigma _{y}=\sigma _{z}=\sigma =\frac{1}{2R^{2}}+\frac{R^{2}\delta ^{2}%
}{8},$ $\xi _{x}=\xi _{y}=\xi _{z}=\xi .$ Then, the effective Hamiltonian
reduce to

\begin{equation}
H=-De^{-\sigma }\cos p+V(R,\xi )+\frac{\Lambda }{2\sqrt{\pi }^{D}R^{D}}
\label{8}
\end{equation}%
where $D=1,2,$ or $3$ reflects the dimension of the system. It can be seen
from Eq.(8), $D$ plays an important role in the effective Hamiltonian.

The group velocity of the wave packet is

\begin{equation}
v_{g}\equiv \frac{\partial H}{\partial p}=\overset{\cdot }{\xi }=\frac{\tan p%
}{m^{\ast }}  \label{9}
\end{equation}%
with the inverse effective mass

\begin{equation}
\frac{1}{m^{\ast }}\equiv \frac{\partial ^{2}H}{\partial p^{2}}=\cos
p\bullet e^{-\sigma }  \label{10}
\end{equation}%
It is important to note that, because this effective mass is related to the
quasi-momentum, the system will exist rich dynamical phenomenons.
Especially, when $\cos p<0$, the effective mass is negative and the system
can exist the localized solution, i.e. soliton solution. For a untilted
trap, $V(R,\xi )=0$, the momentum $p(t)$ remains a constant $p_{0}$.
Initially, we set $\xi _{0}=0$ and $\delta _{0}=0$. In this case, the
effective Hamiltonian becomes

\begin{equation}
H=-D\cos p_{0}\bullet e^{-\sigma }+\frac{\Lambda }{2\sqrt{\pi }^{D}R^{D}}
\label{11}
\end{equation}

Clearly, the dynamical properties of the system not only governed by the
parameter $\Lambda $, but also influenced by the dimension $D$. As discussed
above, the dynamics of the wavepacket are modified significantly by the sign
of $\cos p_{0}$. So two cases with $\cos p_{0}>0$ and $\cos p_{0}<0$ will be
discussed respectively.

\subsection{$\cos p_{0}>0$}

Let us first consider the case with $\cos p_{0}>0$, following this case
there exist two nonlinear phenomenons.

Self-trapping: Equation (9) shows that when the diverging effective mass $%
m^{\ast }\rightarrow \infty $, the group velocity of the wave packet $%
v_{g}\rightarrow 0$, this corresponding to a self-trapping solution. Thus,
for $t\rightarrow \infty $ one has $R\rightarrow R_{\max }$, $\delta
\rightarrow \infty $ and $\overset{\cdot }{\xi }\rightarrow 0,H\rightarrow
\frac{\Lambda }{2\sqrt{\pi }^{D}R_{\max }^{D}}$. \ Because the Hamiltonian
is conserved, $H=H_{0}>0$, so when $H_{0}>0$, we get the maximum value of
the width

\begin{equation}
R_{\max }=(\frac{\Lambda }{\frac{\Lambda }{R_{0}^{D}}-2D\sqrt{\pi }^{D}\cos
p_{0}\bullet e^{-\frac{1}{2R_{0}^{2}}}})^{1/D}  \label{12}
\end{equation}

Diffusion: In this case, for $t\rightarrow \infty $, one has $R\rightarrow
\infty $, $\delta \rightarrow 0$ and $\overset{\cdot }{\xi }\rightarrow
\frac{1}{2}\sin p_{0}$, as well as the effective Hamiltonian $H\rightarrow
-D\cos p_{0}$. So the condition $-\cos p_{0}\leq $ $H_{0}\leq 0$
corresponding to the diffusive region of the system. The critical condition
between self-trapping and diffusion is given by $H_{0}=0$, i.e., $-D\cos
p_{0}\bullet e^{-\frac{1}{2R_{0}^{2}}}+\frac{\Lambda }{2\sqrt{\pi }%
^{D}R_{0}^{D}}=0$, one can get

\begin{equation}
\Lambda _{c}=2\sqrt{\pi }^{D}DR_{0}^{D}\cos p_{0}\bullet e^{-\frac{1}{%
2R_{0}^{2}}}  \label{13}
\end{equation}%
The self-trapping phenomenon occurs at $\Lambda >\Lambda _{c}$, and the
diffusion occurs at $\Lambda <\Lambda _{c}$. By observing the equation (12),
one readily obtain

\begin{equation}
R_{0}/R_{\max }=(1-\frac{\Lambda _{c}}{\Lambda })^{1/D}  \label{14}
\end{equation}%
It is clear that when D=1, Eqs. (13) and (14) reduce to the corresponding
results given in\cite{c11}.

\bigskip

\subsection{$\cos p_{0}<0$}

\bigskip When $\cos p_{0}<0$, the dynamics of the system (11) can be
governed by (see Eq.(7)).

\bigskip
\begin{eqnarray}
\overset{\cdot }{\delta } &=&-D|\cos p_{0}|(\frac{2}{R^{4}}-\frac{\delta ^{2}%
}{2})e^{-\sigma }+\frac{\Lambda }{\sqrt{\pi }^{D}R^{D+2}}  \notag \\
\overset{\cdot }{R} &=&-\frac{R}{2}\delta e^{-\sigma }|\cos p_{0}|
\label{15}
\end{eqnarray}
This system has moving soliton solution. The moving soliton solution
corresponding to the fixed points $\overset{\cdot }{R}=0$ and $\overset{%
\cdot }{\delta }=0$, $\overset{\cdot }{\xi }=$constant. One readily gets
from (15):

\begin{equation}
\delta _{0}=0,\Lambda _{c}=2\sqrt{\pi }^{D}R_{0}^{D-2}|\cos p_{0}|e^{-\frac{1%
}{2R_{0}^{2}}}  \label{16}
\end{equation}%
We now discuss the stability of this soliton state. Let $\delta =\delta
_{0}+\delta ^{\prime },R=R_{0}+R^{\prime },$ and linearize Eq.(15) at the
soliton state, we have:

\begin{equation}
\binom{\overset{\cdot }{\delta ^{\prime }}}{\overset{\cdot }{R^{\prime }}}=%
\frac{\Lambda }{4\sqrt{\pi }^{D}R_{0}^{D+3}}\left(
\begin{array}{cc}
0 & 4(2-D-1/R_{0}^{2}) \\
-R_{0}^{6} & 0%
\end{array}%
\right) \binom{\delta ^{\prime }}{R^{\prime }}  \label{17}
\end{equation}%
one obtains its eigenvalue $\lambda ^{2}=\frac{1}{2}%
R_{0}^{3}(D-2+1/R_{0}^{2}).$ It is clear that when $D=1,$ the soliton state
is a center point (note that $R_{0}>1$), stable soliton solution and
breather solution can exist in $1D$ case. The dynamical behaviors of the
wavepacket in $1D$ case is discussed in detail in Ref.\cite{c11}. When $D>1$%
, however, the soliton solution is a saddle point, no stable soliton and
breather solutions can exist in the system. The numerical results for the
phase diagram in the $R-\delta $ space given by Eq.(15) for $2D$ and $3D$
cases are shown in Fig.1. One can find that soliton and breather cannot
exist in higher dimensional system. When $\Lambda >\Lambda _{c}$, for $%
t\rightarrow \infty ,$ one has $R\rightarrow R_{\min }<R_{0}$ and $\delta
\rightarrow \infty $. It means that the BEC wavepacket remains finite
finally and the self-trapping phenomenon occurs. While $\Lambda <\Lambda
_{c} $, the situation is changed. It is shown that $R\rightarrow \infty $
and $\delta \rightarrow 0$, diffusion takes place. That is, the dynamics of
the BEC wavepackets in $1D$, $2D$ and $3D$ optical lattices are very
different. In $1D$, soliton, breather, and self-trapping states can exist.
In 2D and 3D systems, however, only self-trapping state can exist when $%
\Lambda >\Lambda _{c}$.

\bigskip
\begin{figure}[tbh]
\begin{center}
\rotatebox{0}{\resizebox *{9.0cm}{5.0cm} {\includegraphics
{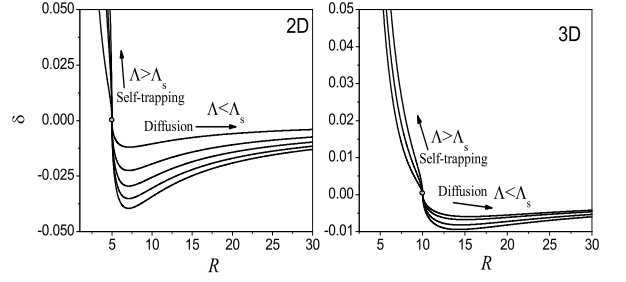}}}
\end{center}
\caption{The Phase diagram of Eq. (15) for $2D$ [(a), $R_{0}=5$, $\protect%
\delta _{0}=0$] and $3D$ case [(b), $R_{0}=10$, $\protect\delta _{0}=0$]}
\end{figure}

\bigskip In this case ($\cos p_{0}<0$), after the self-trapping occurs, $%
R\rightarrow R_{\min }<R_{0}$, $\delta \rightarrow \infty $ and $\overset{%
\cdot }{\xi }\rightarrow 0$, $H\rightarrow \frac{\Lambda }{2\sqrt{\pi }%
^{D}R_{\min }^{D}}$, one can get from $H_{0}=H=-D\cos p_{0}\bullet e^{-\frac{%
1}{2R_{0}^{2}}}+\frac{\Lambda }{2\sqrt{\pi }^{D}R_{0}^{D}}$

\begin{equation}
R_{0}/R_{\min }=(1+\frac{\Lambda ^{\prime }}{\Lambda })^{1/D}  \label{18}
\end{equation}%
where $\Lambda ^{\prime }=2\sqrt{\pi }^{D}DR_{0}^{D}|\cos p_{0}|e^{-\frac{1}{%
2R_{0}^{2}}}$, and $\Lambda >\Lambda _{c}$.

The first and second panels in Fig.2 show the analytical results for the
dynamical diagram of the wavepacket in the $\cos p_{0}-\Lambda _{c}$ plane.
It reveals that the dynamics of the system can be influenced dramatically by
the dimension. Compared with the $1D$ case, the results are modified greatly
in $2D$ and $3D$ systems. The region of the self-trapping decreases quickly
with $D$, while the regions of the diffusion increases with $D$. We can
clearly find that the soliton and breather solutions disappear in $2D$ and $%
3D$ systems. It is also shown that, the effect of $D$ on the regions of the
self-trapping is more significant when the width of wavepacket is increased.

\bigskip
\begin{figure}[tbh]
\begin{center}
\rotatebox{0}{\resizebox *{9.5cm}{9.5cm} {\includegraphics
{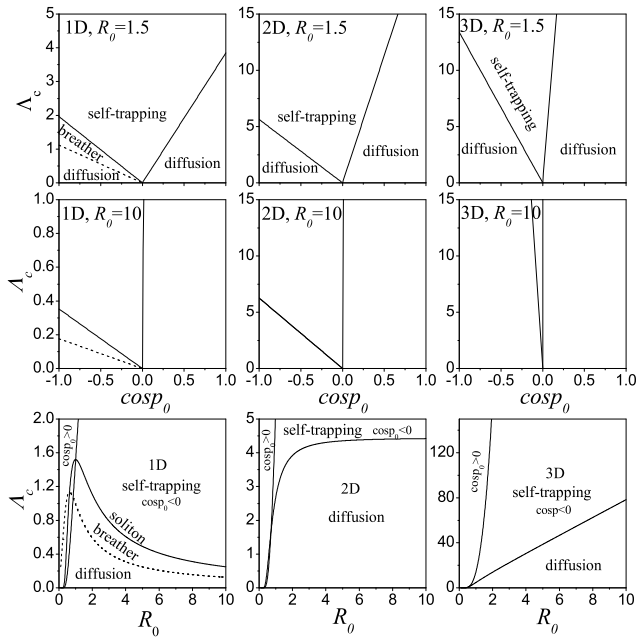}}}
\end{center}
\caption{The critical value $\Lambda _{c}$ against $\cos p_{0}$ (the first
two panels) and initial width $R_{0}$ (the third panel)}
\end{figure}

The critical value $\Lambda _{c}$ against the initial wavepacket width $%
R_{0} $ given by Eqs.(13) and (16) are also described in the third panel in
Fig.2. It is clear that the variation of self-trapping region against $R_{0}$
is very different for $1D$, $2D$, and $3D$ cases, especially, for $\cos
p_{0}<0$ case. When $\cos p_{0}<0$, $\Lambda _{c}$ reaches a maximum in the
beginning and then decreases with increasing $R_{0}$ in $1D$. However, this
relation is modified in $2D$, where $\Lambda _{c}$ increases with $R_{0}$
sharply when $R_{0}<3$ and finally $\Lambda _{c}$ remains an invariable
value when $R_{0}>3$. Especially, $\Lambda _{c}$ increases with $R_{0}$
linearly in $3D$. Moreover, for a fixed $R_{0}$, $\Lambda _{c}$ in $3D$ case
is several times, even one order of magnitude larger than that in $1D$ and $%
2D$. That is, if the initial wavepacket is given, the higher the dimension
is, the greater the atomic interactions or atom number are needed in order
to get into the self-trapping state. Thus it can be seen the dimension has
significant effect on the critical condition for the occurrence of the
self-trapping state.

\begin{figure}[tbh]
\begin{center}
\rotatebox{0}{\resizebox *{9.0cm}{5.0cm} {\includegraphics
{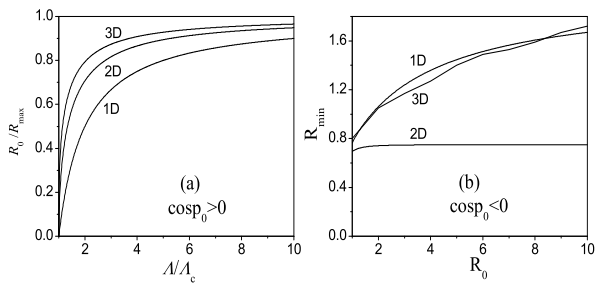}}}
\end{center}
\caption{The value of $R_{0}/R_{\max }$ for $\cos p_{0}>0$ (a) and the
minimum value $R_{\min }$ against the initial width $R_{0}$ for $\cos
p_{0}<0 $ (b)}
\end{figure}

Fig.3(a) shows the ratio $R_{0}/R_{\max }$ versus ratio $\Lambda /\Lambda
_{c}$ for $\cos p_{0}>0$. When the self-trapping occurs, one can find that
the final wavepacket width $R_{\max }$ is larger than the initial wavepacket
width $R_{0}$. For a fixed $\Lambda /\Lambda _{c}$, $R_{0}/R_{\max }$
increases with increasing $D$. It means that, when $\Lambda >\Lambda _{c}$,
the BEC cloud can easily get into self-trapping state with little adjustment
of the initial wavepacket for higher dimension case. In addition, it is
clearly that $R_{0}/R_{\max }$ increases with $\Lambda /\Lambda _{c}$ and
close to 1 eventually for $1D$, $2D$, and $3D$ cases. In other words, for
sufficiently large $\Lambda $, the BEC cloud will quickly get into
self-trapping state for all the case and the effect of dimension is weakened.

Fig.3(b) shows the minimum wavepacket width $R_{\min }$ against the initial
wavepacket width $R_{0}$ for $\cos p_{0}<0$. As is expressed, $R_{\min }$ is
always smaller than $R_{0}$. More interestingly, $R_{\min }$ increases with $%
R_{0}$ in $1D$ and $3D$, while $R_{\min }$ keeps a constant ($0.75$) and
nearly independent with $R_{0}$. On the other hand, because the effect mass
given by Eq.(10) is negative, the BEC cloud will finally squeeze into
several lattice points for $1D$ and $3D$ case and about two lattice points
for $2D$ case.

\begin{figure}[tbh]
\begin{center}
\rotatebox{0}{\resizebox *{10.0cm}{10.0cm} {\includegraphics
{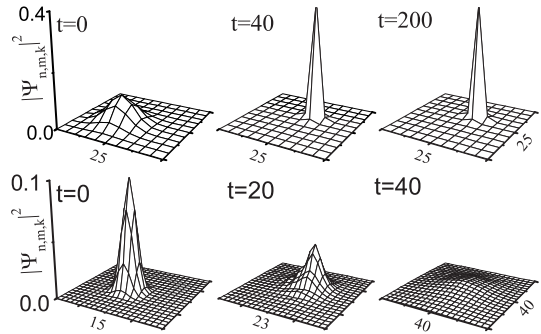}}}
\end{center}
\caption{The dynamics of the $2D$ wavepacket with $\Lambda =6.5>\Lambda
_{c}\simeq 4$ (the first panel) and $\Lambda =3.5<\Lambda _{c}\simeq 4$ (the
second panel), $R_{0}=2.5$, $p_{0}=3\protect\pi /4$}
\end{figure}

\section{\protect\bigskip NUMERICAL SIMULATIONS ON THE GPE}

To confirm the above theoretical predictions, the direct numerical
simulations of the full discrete GPE(5) are presented in Figs.4-9.

The numerical results for the dynamics of $2D$ wavepacket with small initial
width ($R_{0}=2.5$) are provided in Fig.4. It is explicit that when $\Lambda
>\Lambda _{c}$ (first panel) the BEC cloud is quickly squeezed into two
optical lattices finally and the width of the wavepacket remains finite
(self-trapping). On the contrary, when $\Lambda <\Lambda _{c}$ (second
panel), as it is expected the wavepacket will expanding indefinitely with
time (diffusion).

Figs.5-7 show the numerical results of the wavepacket with wider initial
wavepacket ($R_{0}=10$). The results with $\Lambda =5>\Lambda _{c}\approx
4.4 $ are shown in Fig.5. One can find that the wavepacket is also trapped
into several lattices (at $t\approx 100$) as wavepacket moving. In this
case, however, the self-trapping occurs only during a certain time ($%
t\lesssim 100$, this time scale decreasing with $\Lambda $) but then the
wavepacket start to diffuse. If we strengthen the nonlinear term to $\Lambda
=20>>\Lambda _{c}=4.4$, the associated results are shown in Fig.6. It
indicates that the condensate is trapped almost in two optical lattices
eventually and diffusion is not occur. Also, Fig.7 provides the results with
$\Lambda <\Lambda _{c}$, as expected the diffusion result is obtained.

\begin{figure}[tbh]
\begin{center}
\rotatebox{0}{\resizebox *{8.5cm}{8.5cm} {\includegraphics
{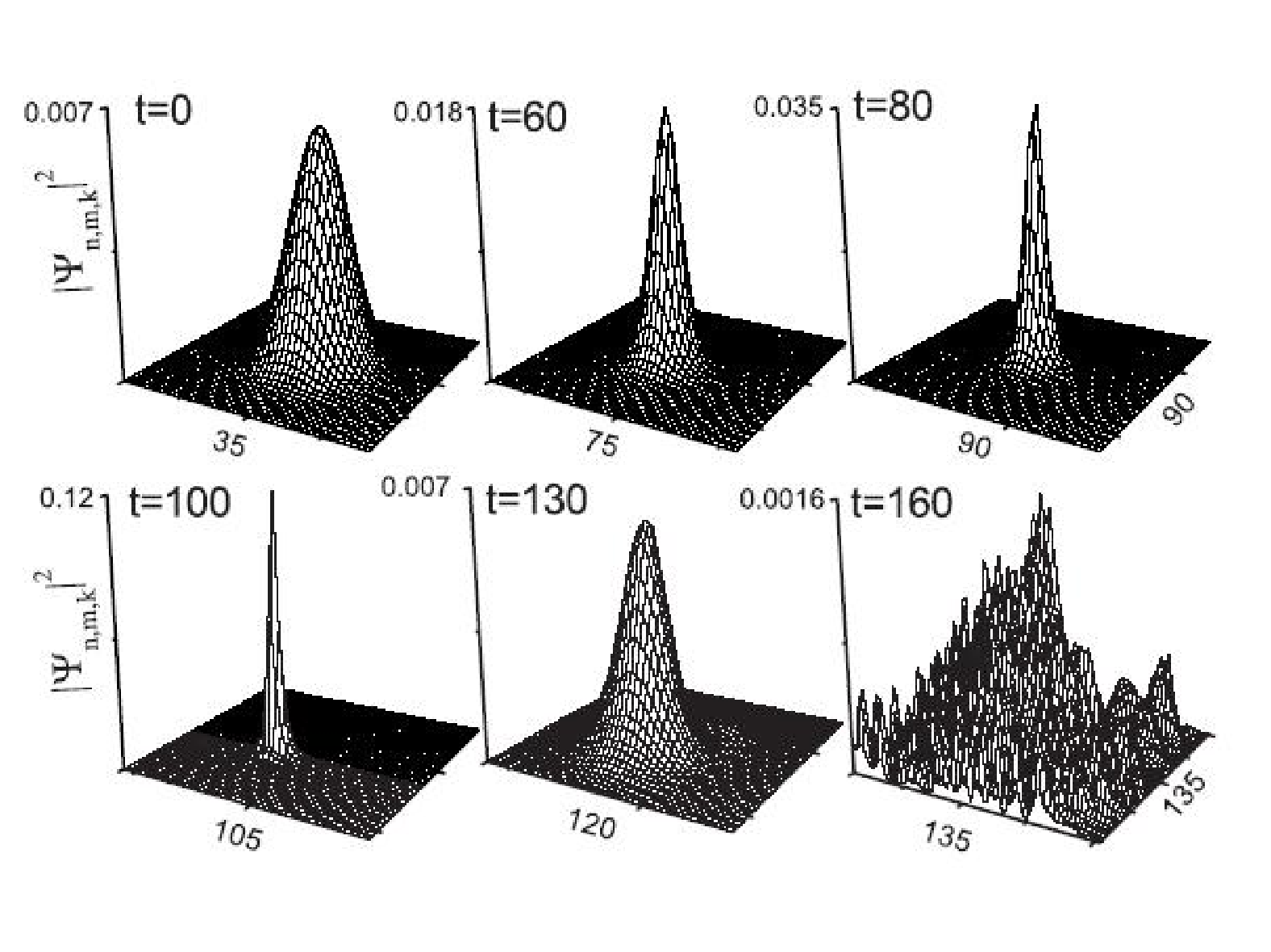}}}
\end{center}
\caption{The dynamics of the $2D$ wavepacket with $\Lambda =5>\Lambda
_{c}\simeq 4.4$, $R_{0}=10$, $p_{0}=3\protect\pi /4$}
\end{figure}

\bigskip
\begin{figure}[tbh]
\begin{center}
\rotatebox{0}{\resizebox *{8.0cm}{7.0cm} {\includegraphics
{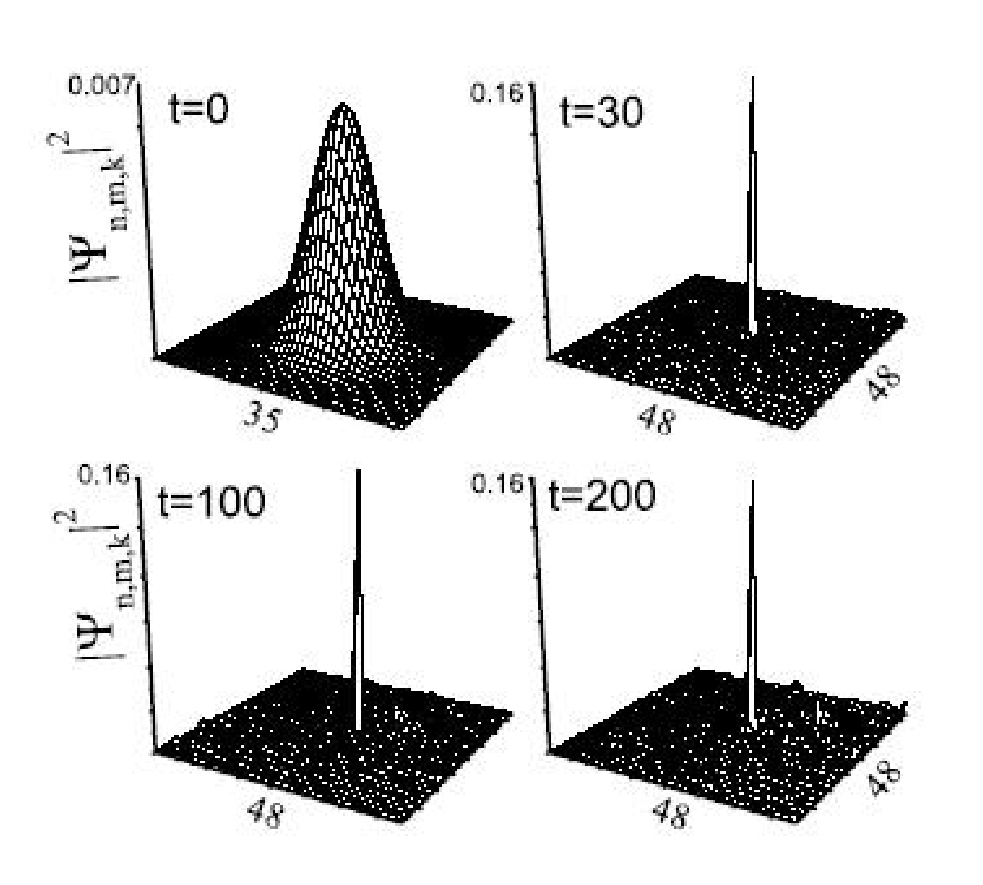}}}
\end{center}
\caption{The dynamics of the $2D$ wavepacket with $\Lambda =20>>\Lambda
_{c}\simeq 4.4$, $R_{0}=10$, $p_{0}=3\protect\pi /4$}
\end{figure}

\bigskip
\begin{figure}[tbh]
\begin{center}
\rotatebox{0}{\resizebox *{8.0cm}{7.0cm} {\includegraphics
{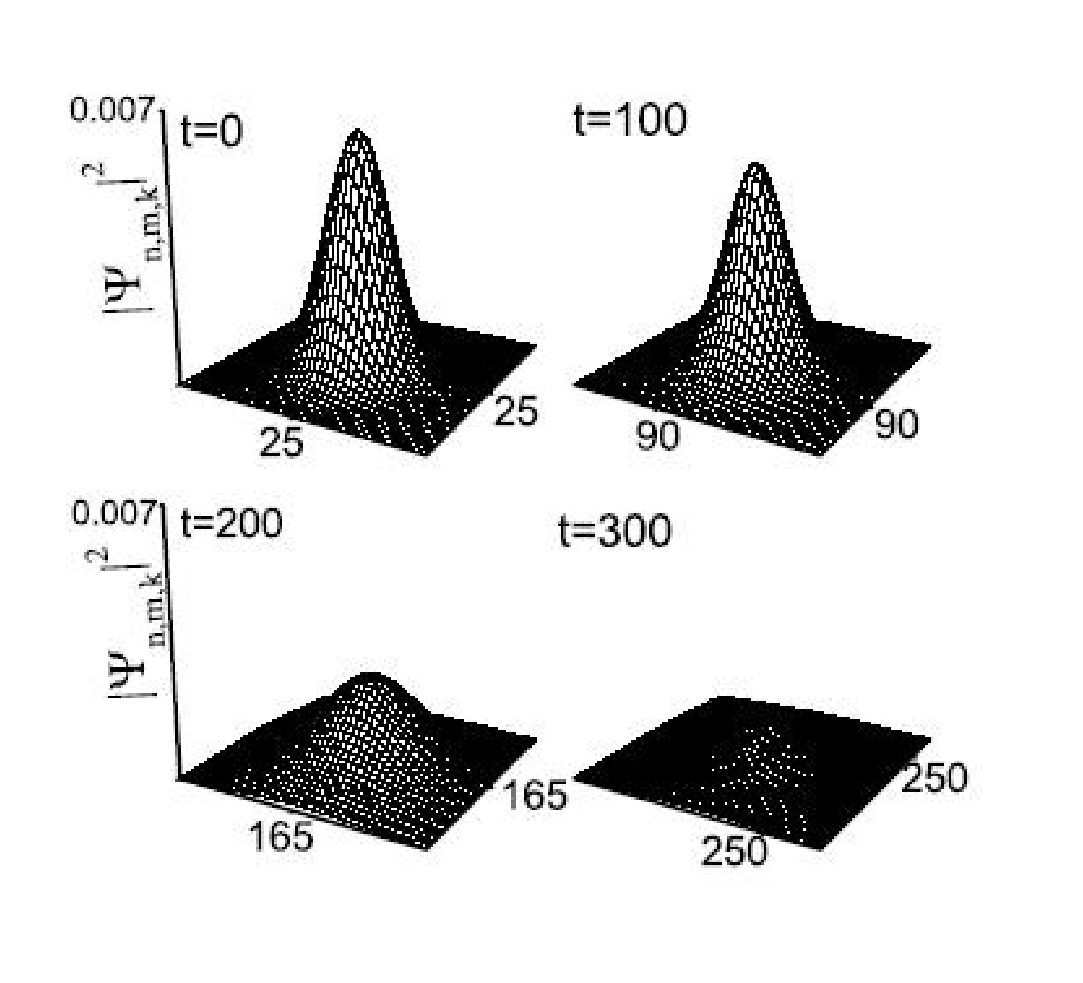}}}
\end{center}
\caption{The dynamics of the $2D$ wavepacket with $\Lambda =3<\Lambda
_{c}\simeq 4.4$, $R_{0}=10$, $p_{0}=3\protect\pi /4$}
\end{figure}

\bigskip Figs.8-9 show the numerical results for the dynamics of 3D
wavepacket. Differ from the results in $2D$, when $\Lambda >\Lambda _{c}$
(the first panel in Fig.8 and Fig.9, respectively), the $3D$ wavepacket gets
into self-trapping state and finally, the wavepacket splits into two pieces
and localized state appears. The results with $\Lambda <\Lambda _{c}$ (the
second panel in Fig.8 and Fig.9, respectively) give the diffusion results.
It is clearly that the above numerical results are in good agreement with
our variational predications. It can be seen the dimension of the lattices
has significant effect on the dynamic of BEC wavepackets moving in optical
lattices.

\begin{figure}[tbh]
\begin{center}
\rotatebox{0}{\resizebox *{8.0cm}{6.0cm} {\includegraphics
{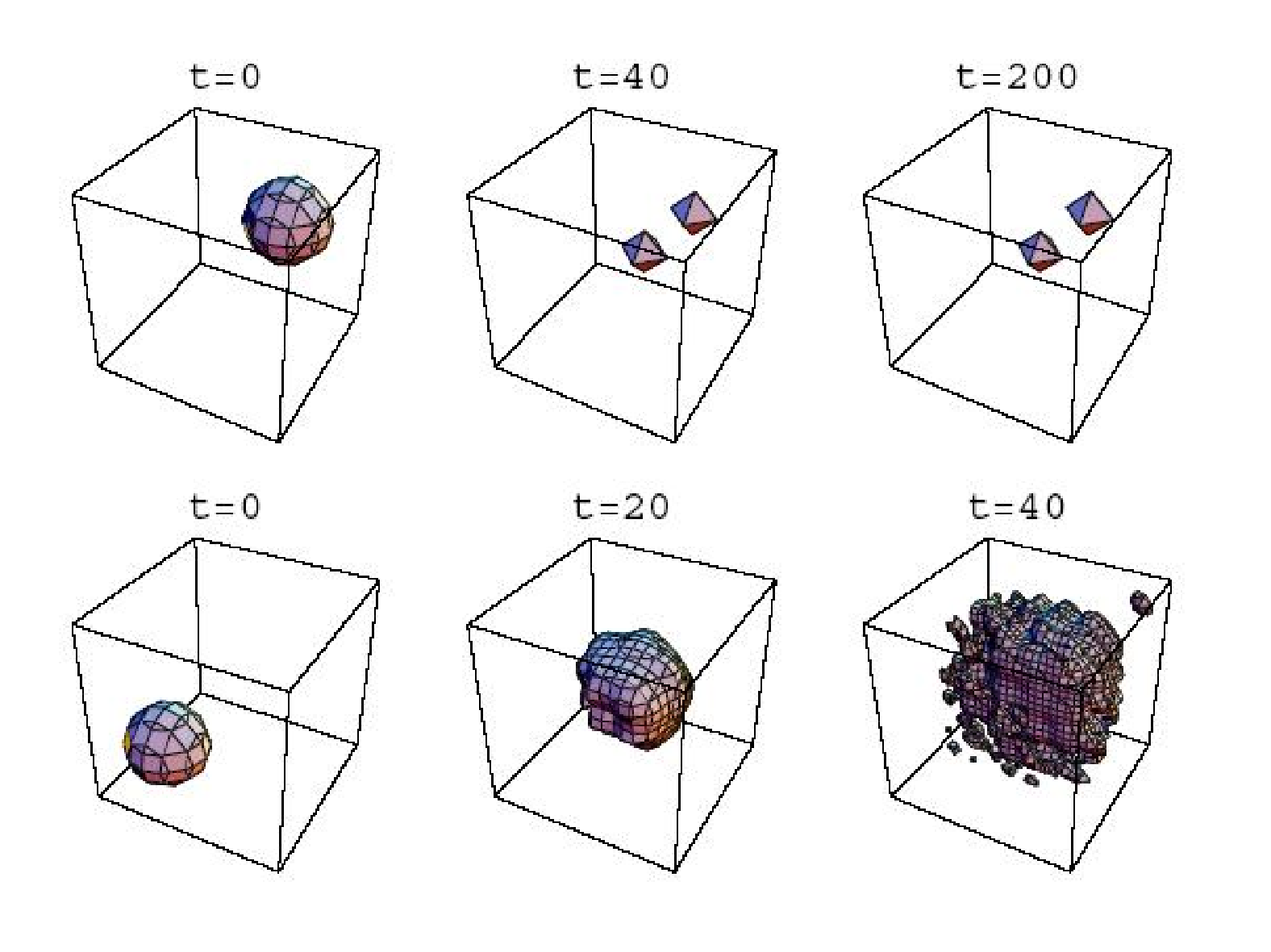}}}
\end{center}
\caption{The dynamics of the $3D$ wavepacket with $\Lambda =30>\Lambda _{c}$
$\simeq 18$ (the first panel) and $\Lambda =15<\Lambda _{c}$ $\simeq 18$
(the second panel), $R_{0}=2.5$, $p_{0}=3\protect\pi /4$}
\end{figure}

\bigskip
\begin{figure}[tbh]
\begin{center}
\rotatebox{0}{\resizebox *{8.0cm}{6.0cm} {\includegraphics
{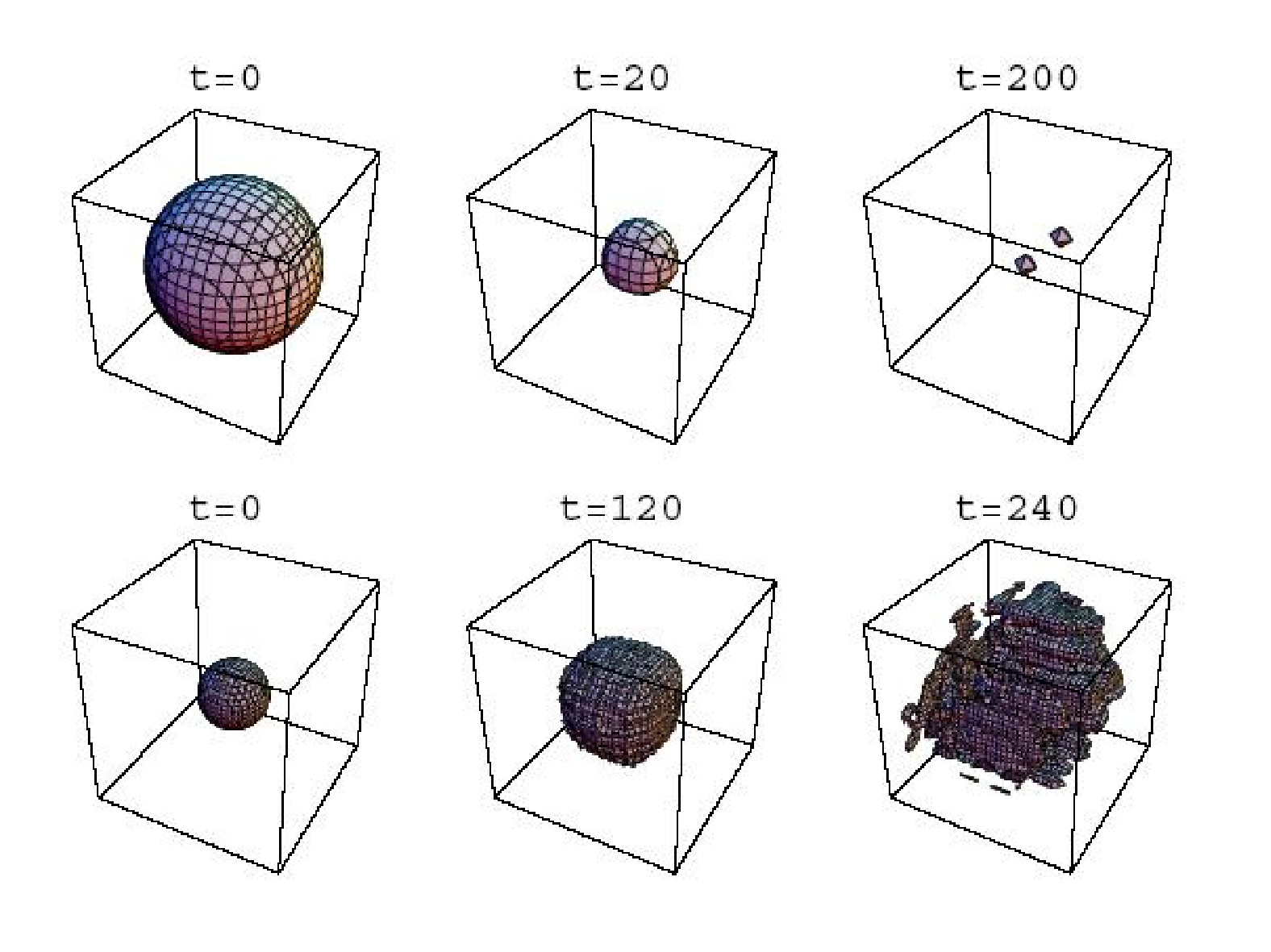}}}
\end{center}
\caption{The dynamics of the $3D$ wavepacket with $\Lambda =90>\Lambda _{c}$
$\simeq 79$ (the first panel) and $\Lambda =50<\Lambda _{c}$ $\simeq 79$
(the second panel), $R_{0}=10$, $p_{0}=3\protect\pi /4$}
\end{figure}

\section{\protect\bigskip Conclusion}

To summarize, we have investigated the dynamics of a BEC wavepacket
loaded into a deep multi-dimensional periodic optical lattices both
numerically and analytically, focusing on the effect of the lattice
dimension. Our study shows that  the dynamics of the BEC wavepacket
is quite different in $1D$, $2D$ or  $3D$ systems. For example, the
stable moving soliton and breather states that exist in $1D$ lattice
system no longer stand in $2D$ and $3D$ lattice systems. We also
obtain an analytical criterion for the self-trapping state of BEC
and  phase diagram for self-trapping, soliton, breather, or
diffusion of the BEC cloud for 1D, 2D and 3D, respectively. The
above  analytical results are confirmed by our directly solving the
GPE. We hope that our studies will stimulate the experiments in the
direction.

\section*{Acknowledgments}

This work is supported by the National Natural Science Foundation of
China under Grant No. 10474008, 10604009, 10475066, the National
Fundamental Research Programme of China under Grant No.
2006CB921400, the Natural Science Foundation of Gansu province under
Grant No. 3ZS051-A25-013, and by NWNU-KJCXGC-03-17.

\end{document}